\documentclass[12pt]{article}
\usepackage{a4wide,epsfig,psfrag,amsmath,amssymb,cite,scalefnt,axodraw4j,pstricks,caption,subcaption}

\usepackage{color}

\newcommand{\gsim}{\;\rlap{\lower 3.5 pt \hbox{$\mathchar \sim$}} \raise 1pt
 \hbox {$>$}\;}
\newcommand{\lsim}{\;\rlap{\lower 3.5 pt \hbox{$\mathchar \sim$}} \raise 1pt
 \hbox {$<$}\;}

\newcommand{\nfour}{{\cal N}=4}

\begin{document}

\title{\vskip-3cm{\baselineskip14pt
    \begin{flushleft}
      \normalsize SFB/CPP-13-42\\
      \normalsize TTP13-23
  \end{flushleft}}
  \vskip1.5cm
  Two-loop static potential in $\nfour$ supersymmetric Yang-Mills theory
}
\author{\small 
  Mario Prausa and Matthias Steinhauser
  \\[1em]
  {\small\it Institut f{\"u}r Theoretische Teilchenphysik}\\
  {\small\it Karlsruhe Institute of Technology (KIT)}\\
  {\small\it 76128 Karlsruhe, Germany}
}

\date{}

\maketitle

\thispagestyle{empty}

\begin{abstract}

  We compute the soft contribution to the static energy of two heavy colour
  sources interacting via a $\nfour$ supersymmetric Yang-Mills theory.  Both
  singlet and octet colour configurations are considered.  Our
  calculations complete recent considerations of the ultrasoft contributions.

\medskip

\noindent
PACS numbers: 11.30.Pb, 12.38.Bx, 12.39.Hg, 12.60.Jv

\end{abstract}

\thispagestyle{empty}



\section{Introduction}

Although $\nfour$ supersymmetric Yang-Mills (SYM) theory is not realized in nature
it has received increasing attraction in the recent years.  The main reason
for this is the conjecture of a duality between $\nfour$ SYM and a certain
class of string theories which is usually called the AdS/CFT
correspondence~\cite{Maldacena:1997re}. To test the correspondence it would be
ideal to have all-order perturbative results which can be evaluated for large
values of the coupling and then be compared to string theory calculations.
However, in general only a few terms can be computed in the perturbative
expansion and in the strong-coupling limit. One hopes to obtain information
about the AdS/CFT correspondence from their comparison.

Among the interesting quantities which one can consider there is the
static energy of two infinitely-heavy colour sources in the fundamental
representation of $SU(N_c)$. It has been considered in the strong and weak
coupling limit in Refs.~\cite{Rey:1998ik,Maldacena:1998im}
and~\cite{Erickson:1999qv,Erickson:2000af}, respectively.  A 
systematic one-loop calculation in a framework analogue to the one applied in
QCD has been performed in Ref.~\cite{Pineda:2007kz}. It has been noted that
already at this loop-order ultrasoft contributions have to be taken into
account which is due to the massless scalar particles present in $\nfour$ SYM.

Recently, in Ref.~\cite{Stahlhofen:2012zx} the two-loop ultrasoft
contribution to the static energy has been computed. On its own 
it still contains poles in $\epsilon$ which have been
subtracted using the $\overline{\rm MS}$
scheme. In this paper we provide the soft contribution to the static energy
both for singlet and octet colour configuration which
combines with the ultrasoft result to a finite physical expression for
the static energy. 

In QCD, ultrasoft contributions arise for the first time at three-loop
order~\cite{Appelquist:1977es,Brambilla:1999qa,Kniehl:1999ud,Kniehl:2002br}
since the real radiation of gluons from ultrasoft quarks is suppressed by
$v\sqrt{\alpha_s}$ where $v$ is the velocity of the heavy quark. Thus, after
taking into account the scaling rule $v \sim \alpha_s$ the combination of
emission and absorption process scales like $\alpha_s^3$.

In the next Section we provide some details to our calculation and 
the results are presented in Section~\ref{sec::res}.


\section{Calculation}

The theoretical framework convenient for the computation of the static energy
within $\nfour$ SYM has already been presented in
Refs.~\cite{Pineda:2007kz,Stahlhofen:2012zx}.  Let us for convenience repeat
the main steps which are important for our calculation.

The Lagrange density for $\nfour$ SYM theory reads
\begin{eqnarray}
  {\cal L}_{\nfour} &=&
  -\frac14 F_{\mu\nu}^a F^{\mu\nu\;a} 
  + \frac12 \sum\limits_{i=1}^6 \left(D_\mu \Phi_i\right)^a\left(D^\mu\Phi_i\right)^a
  - \frac i2 \sum\limits_{j=1}^4 \bar\Psi_j^a \gamma_\mu \left(D^\mu \Psi_j\right)^a
  + \dots
  \,,
\end{eqnarray}
where $\Phi_i$ ($i=1,\ldots,6$) represent six (pseudo) scalar particles
and $\Psi_j$ ($j=1,\ldots,4$) four Majorana fermions in the adjoint
representation of $SU(N_c)$, just like the gluon fields $A_\mu$ present in the field
strength tensor $F^{\mu\nu}$.

Following~\cite{Drukker:1999zq} we introduce the Wilson loop
\begin{eqnarray}
  W_C &=& \frac{1}{N_c}\mbox{Tr}{\cal P}{\rm exp}\left[-ig \oint_C{\rm d}\tau
    \left(A_\mu \dot{x}^\mu + \Phi_n|\dot{x}|\right)
  \right]
  \,.
  \label{eq::WC}
\end{eqnarray}
The static energy is obtained by considering
a rectangular path $C$ and taking the limit of large temporal extension
of the expression $(i/T)\ln\langle W_\Box \rangle$~\cite{Pineda:2007kz}.

The interaction with the static colour sources $\psi$ and $\chi$ in the
fundamental representation of $SU(N_c)$ is described via the Lagrange density
\begin{eqnarray}
  {\cal L}_{\rm stat} &=&
  \psi^\dagger\left(i\partial_0-gA_0-g\Phi_n\right)\psi 
  + \chi_c^\dagger\left(i\partial_0 + gA_0^T -g\Phi_n^T\right)\chi_c
  \,.
  \label{eq::Lstat}
\end{eqnarray}

Our aim is the computation of the static energy between two static sources,
one in the fundamental and one in the anti-fundamental representation, to
two-loop accuracy.  It can be written as
\begin{eqnarray}
  E_{s,o} &=& V_{s,o} + \delta E^{\rm US}_{s,o}
  \,,
  \label{eq::Eso}
\end{eqnarray}
where the subscripts ``s'' and ``o'' represent the singlet and octet
representation of the source-anti-source system, respectively.  The one-loop
results for $V_s$, $V_o$ and $\delta E^{\rm US}_{s}$ have been obtained in
Ref.~\cite{Pineda:2007kz} and the one- and two-loop results for $\delta E^{\rm
  US}_{s}$ and $\delta E^{\rm US}_{o}$ have been computed in
Ref.~\cite{Stahlhofen:2012zx}.  In this paper we complete the 
next-to-next-to-leading order calculation
and compute $V_s$ and $V_o$ to two loops. Furthermore, we add a two-loop
diagram to the expression for $\delta E^{\rm US}_{o}$
in~\cite{Stahlhofen:2012zx} which was omitted in that reference.

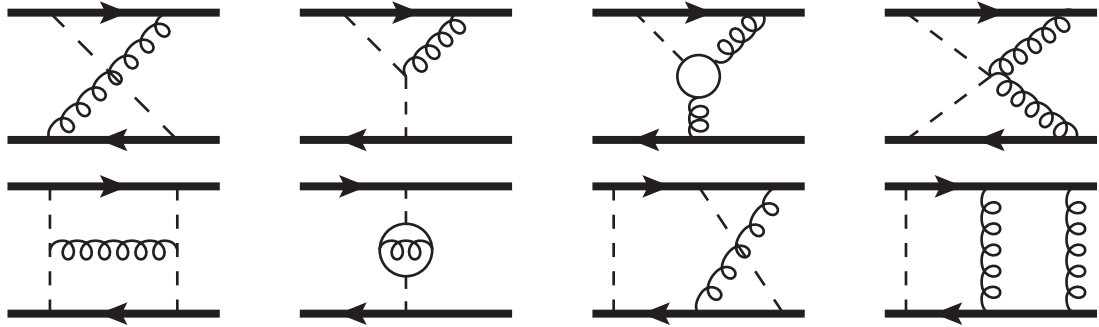
\begin{figure}[t]
  \centering
  \begin{tabular}{cccc}
    \fcolorbox{white}{white}{
      \begin{picture}(84,58) (14,-27)
        \SetWidth{3.0}
        \SetColor{Black}
        \Line[arrow,arrowpos=0.5,arrowlength=8.333,arrowwidth=3.333,arrowinset=0.2](16,26)(96,26)
        \Line[arrow,arrowpos=0.5,arrowlength=8.333,arrowwidth=3.333,arrowinset=0.2](96,-22)(16,-22)
        \SetWidth{1.0}
        \Line[dash,dashsize=7](32,26)(80,-22)
        \Gluon(32,-22)(80,26){3.5}{7}
      \end{picture}
    } &
    \fcolorbox{white}{white}{
      \begin{picture}(84,58) (14,-27)
        \SetWidth{3.0}
        \SetColor{Black}
        \Line[arrow,arrowpos=0.5,arrowlength=8.333,arrowwidth=3.333,arrowinset=0.2](16,26)(96,26)
        \Line[arrow,arrowpos=0.75,arrowlength=8.333,arrowwidth=3.333,arrowinset=0.2](96,-22)(16,-22)
        \SetWidth{1.0}
        \Line[dash,dashsize=7](32,26)(56,2)
        \Gluon(56,2)(80,26){3.5}{4}
        \Line[dash,dashsize=6](56,2)(56,-22)
      \end{picture}
    } &
    \fcolorbox{white}{white}{
      \begin{picture}(84,58) (14,-27)
        \SetWidth{3.0}
        \SetColor{Black}
        \Line[arrow,arrowpos=0.5,arrowlength=8.333,arrowwidth=3.333,arrowinset=0.2](16,26)(96,26)
        \Line[arrow,arrowpos=0.75,arrowlength=8.333,arrowwidth=3.333,arrowinset=0.2](96,-22)(16,-22)
        \SetWidth{1.0}
        \Arc(56,2)(8,180,540)
        \Line[dash,dashsize=6](32,26)(50,8)
        \Gluon(80,26)(62,8){3.5}{3}
        \Gluon(56,-22)(56,-6){3.5}{2}
      \end{picture}
    } &
    \fcolorbox{white}{white}{
      \begin{picture}(84,58) (14,-27)
        \SetWidth{3.0}
        \SetColor{Black}
        \Line[arrow,arrowpos=0.5,arrowlength=8.333,arrowwidth=3.333,arrowinset=0.2](16,26)(96,26)
        \Line[arrow,arrowpos=0.5,arrowlength=8.333,arrowwidth=3.333,arrowinset=0.2](96,-22)(16,-22)
        \SetWidth{1.0}
        \Line[dash,dashsize=6](24,26)(56,2)
        \Line[dash,dashsize=6](24,-22)(56,2)
        \Gluon(56,2)(88,26){3.5}{5}
        \Gluon(56,2)(88,-22){3.5}{5}
      \end{picture}
    } \\
    \fcolorbox{white}{white}{
      \begin{picture}(84,58) (14,-27)
        \SetWidth{3.0}
        \SetColor{Black}
        \Line[arrow,arrowpos=0.5,arrowlength=8.333,arrowwidth=3.333,arrowinset=0.2](16,26)(96,26)
        \Line[arrow,arrowpos=0.5,arrowlength=8.333,arrowwidth=3.333,arrowinset=0.2](96,-22)(16,-22)
        \SetWidth{1.0}
        \Line[dash,dashsize=6](32,26)(32,-22)
        \Line[dash,dashsize=6](80,-22)(80,26)
        \Gluon(32,2)(80,2){3.5}{6}
      \end{picture}
    } &
    \fcolorbox{white}{white}{
      \begin{picture}(84,58) (14,-27)
        \SetWidth{3.0}
        \SetColor{Black}
        \Line[arrow,arrowpos=0.25,arrowlength=8.333,arrowwidth=3.333,arrowinset=0.2](16,26)(96,26)
        \Line[arrow,arrowpos=0.75,arrowlength=8.333,arrowwidth=3.333,arrowinset=0.2](96,-22)(16,-22)
        \SetWidth{1.0}
        \Arc(56,2)(9.899,135,495)
        \Gluon(46,2)(66,2){3.5}{2}
        \Line[dash,dashsize=6](56,26)(56,12)
        \Line[dash,dashsize=6](56,-8)(56,-22)
      \end{picture}
    } &
    \fcolorbox{white}{white}{
      \begin{picture}(84,58) (14,-27)
        \SetWidth{3.0}
        \SetColor{Black}
        \Line[arrow,arrowpos=0.3,arrowlength=8.333,arrowwidth=3.333,arrowinset=0.2](16,26)(96,26)
        \Line[arrow,arrowpos=0.7,arrowlength=8.333,arrowwidth=3.333,arrowinset=0.2](96,-22)(16,-22)
        \SetWidth{1.0}
        \Line[dash,dashsize=6](24,26)(24,-22)
        \Line[dash,dashsize=6](56,26)(88,-22)
        \Gluon(56,-22)(88,26){3.5}{5}
      \end{picture}
    } &
    \fcolorbox{white}{white}{
      \begin{picture}(84,58) (14,-27)
        \SetWidth{3.0}
        \SetColor{Black}
        \Line[arrow,arrowpos=0.3,arrowlength=8.333,arrowwidth=3.333,arrowinset=0.2](16,26)(96,26)
        \Line[arrow,arrowpos=0.7,arrowlength=8.333,arrowwidth=3.333,arrowinset=0.2](96,-22)(16,-22)
        \SetWidth{1.0}
        \Line[dash,dashsize=6](24,26)(24,-22)
        \Gluon(56,-22)(56,26){3.5}{5}
        \Gluon(88,-22)(88,26){3.5}{5}
      \end{picture}
    }
  \end{tabular}
  \caption[]{\label{fig::dia} One- and two-loop Feynman diagrams contributing
    to $V_s$ and $V_o$. Thick solid lines represent heavy colour sources, thin
    solid lines massless fermions, curled lines gluons, and dashed lines
    massless scalar particles.}
\end{figure}

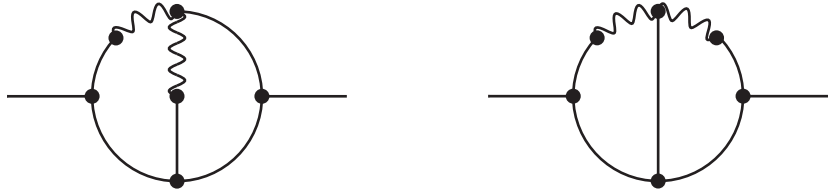
\begin{figure}[t]
  \centering
  \begin{subfigure}[b]{0.4\textwidth}
    \centering
    \fcolorbox{white}{white}{
      \begin{picture}(130,72) (15,-28)
        \SetWidth{1.0}
        \SetColor{Black}
        \Arc(80,8)(32,90,450)
        \Line(80,8)(80,-24)
        \Vertex(80,-24){2.828}
        \SetColor{White}
        \CBox(50,30)(80,43){White}{White}
        \SetColor{Black}
        \Vertex(48,8){2.828}
        \Line(16,8)(48,8)
        \Photon(80,40)(80,8){3}{4}
        \Vertex(80,8){2.828}
        \Vertex(80,40){2.828}
        \Vertex(112,8){2.828}
        \Line(112,8)(144,8)
        \PhotonArc[clock](77.19,15.012)(25.146,143.412,83.585){3}{3}
        \Vertex(57,30){2.828}
      \end{picture}
    }
  \end{subfigure}%
  \begin{subfigure}[b]{0.4\textwidth}
    \centering
    \fcolorbox{white}{white}{
      \begin{picture}(130,73) (15,-27)
        \SetWidth{1.0}
        \SetColor{Black}
        \Arc(80,9)(32,90,450)
        \Vertex(80,-23){2.828}
        \SetColor{White}
        \CBox(43,31)(121,45){White}{White}
        \SetColor{Black}
        \Vertex(48,9){2.828}
        \Line(16,9)(48,9)
        \Vertex(80,41){2.828}
        \Vertex(112,9){2.828}
        \Line(112,9)(144,9)
        \PhotonArc[clock](79.5,10.7)(30.304,137.943,42.057){3}{6}
        \Vertex(102,31){2.828}
        \Line(80,41)(80,-23)
        \Vertex(57,31){2.828}
      \end{picture}
    }
 \end{subfigure}
 \caption{\label{fig::dia2l}
    Families of scalar two-loop Feynman integrals. Solid and wavy
    lines represent relativistic massless and static propagators,
    respectively.}
\end{figure}

We perform the calculation of $V_s$ and $V_o$ in momentum space, in close
analogy to the calculations performed in the context of QCD which are
discussed in detail in the
literature~\cite{Peter:1996ig,Peter:1997me,Schroder:1998vy,Schroder:1999sg,Kniehl:2004rk,Collet:2011kq,Smirnov:2008pn,Smirnov:2009fh,Anzai:2009tm}.
The potential is obtained from the one-particle-irreducible contributions to
four-point functions with momentum exchange $\vec{q}$ between the static
sources.  Sample Feynman diagrams are shown in Fig.~\ref{fig::dia}.  For the
singlet contribution only non-abelian contributions have to be considered.  As
a consequence there are no contributions which contain so-called
pinch-singularities of the form
\begin{eqnarray}
  \frac{1}{(k_0 + i0)(k_0 - i0)}
  \,,
  \label{eq::pinch}
\end{eqnarray}
where $k_0$ is the 0-component of the loop momentum and thus all two-loop integrals
can be reduced to one of the families shown in Fig.~\ref{fig::dia2l}.  We
perform the reduction of the scalar integrals to master integrals with the
help of {\tt FIRE}~\cite{FIRE}.  The analytic results for the master
integrals are taken from Ref.~\cite{Smirnov:2003kc}.

The colour-octet potential needs special attention since Feynman diagrams with
pinch contributions (cf. Eq.~(\ref{eq::pinch})) contribute to $V_o$.
The corresponding integrals cannot be computed directly but can be 
reduced to integrals without pinches using the methods described in
Refs.~\cite{Kniehl:2004rk,Collet:2011kq}. In our calculation we have exploited
the exponentiation of the colour-singlet potential in order to establish
relations between Feynman integrals with the same colour factor. As an example
let us consider the ladder-type diagrams which have colour factors
$C_F^3$, $C_F^2C_A$ and $C_FC_A^2$. Feynman diagrams with pinches are only
present in the first two cases. Exponentiation requires that the sum
of all contributions proportional to $C_F^3$ or $C_F^2C_A$
vanish which can be expressed through the following graphical
equations\footnote{For simplicity we only consider ladder-type diagrams.}
\begin{eqnarray*}
  \begin{array}{c}
    \scalebox{0.75}{\fcolorbox{white}{white}{
      \begin{picture}(12,20) (34,-14)
        \SetWidth{2.5}
        \SetColor{Black}
        \Line(28,4)(52,4)
        \Line(52,-12)(28,-12)
        \SetWidth{1.0}
        \Line(32,4)(32,-12)
        \Line(48,4)(48,-12)
        \Line(40,-12)(40,4)
      \end{picture}
    }}
  \end{array} +
  \begin{array}{c}
    \scalebox{0.75}{\fcolorbox{white}{white}{
      \begin{picture}(12,20) (34,-14)
        \SetWidth{2.5}
        \SetColor{Black}
        \Line(28,4)(52,4)
        \Line(52,-12)(28,-12)
        \SetWidth{1.0}
        \Line(32,4)(32,-12)
        \Line(40,4)(48,-12)
        \Line(40,-12)(48,4)
      \end{picture}
    }}
  \end{array} + 
  \begin{array}{c}
    \scalebox{0.75}{\fcolorbox{white}{white}{
      \begin{picture}(12,20) (34,-14)
        \SetWidth{2.5}
        \SetColor{Black}
        \Line(28,4)(52,4)
        \Line(52,-12)(28,-12)
        \SetWidth{1.0}
        \Line(40,4)(32,-12)
        \Line(48,4)(48,-12)
        \Line(40,-12)(32,4)
      \end{picture}
    }}
  \end{array} + 
  \begin{array}{c}
    \scalebox{0.75}{\fcolorbox{white}{white}{
      \begin{picture}(12,20) (34,-14)
        \SetWidth{2.5}
        \SetColor{Black}
        \Line(28,4)(52,4)
        \Line(52,-12)(28,-12)
        \SetWidth{1.0}
        \Line(48,4)(32,-12)
        \Line(32,4)(48,-12)
        \Line(40,-12)(40,4)
      \end{picture}
    }}
  \end{array} +
  \begin{array}{c}
    \scalebox{0.75}{\fcolorbox{white}{white}{
      \begin{picture}(12,20) (34,-14)
        \SetWidth{2.5}
        \SetColor{Black}
        \Line(28,4)(52,4)
        \Line(52,-12)(28,-12)
        \SetWidth{1.0}
        \Line(40,4)(32,-12)
        \Line(32,4)(48,-12)
        \Line(40,-12)(48,4)
      \end{picture}
    }}
  \end{array} +
  \begin{array}{c}
    \scalebox{0.75}{\fcolorbox{white}{white}{
      \begin{picture}(12,20) (34,-14)
        \SetWidth{2.5}
        \SetColor{Black}
        \Line(28,4)(52,4)
        \Line(52,-12)(28,-12)
        \SetWidth{1.0}
        \Line(48,4)(32,-12)
        \Line(40,4)(48,-12)
        \Line(40,-12)(32,4)
      \end{picture}
    }}
  \end{array} 
  + \text{\parbox{5em}{\centering (iteration terms)}} &=& 0   
  \,,\\
  \frac12\left(
  \begin{array}{c}
    \scalebox{0.75}{\fcolorbox{white}{white}{
      \begin{picture}(12,20) (34,-14)
        \SetWidth{2.5}
        \SetColor{Black}
        \Line(28,4)(52,4)
        \Line(52,-12)(28,-12)
        \SetWidth{1.0}
        \Line(32,4)(32,-12)
        \Line(40,4)(48,-12)
        \Line(40,-12)(48,4)
      \end{picture}
    }}
  \end{array} +
  \begin{array}{c}
    \scalebox{0.75}{\fcolorbox{white}{white}{
      \begin{picture}(12,20) (34,-14)
        \SetWidth{2.5}
        \SetColor{Black}
        \Line(28,4)(52,4)
        \Line(52,-12)(28,-12)
        \SetWidth{1.0}
        \Line(40,4)(32,-12)
        \Line(48,4)(48,-12)
        \Line(40,-12)(32,4)
      \end{picture}
    }}
  \end{array} \right) + \frac32 
  \begin{array}{c}
    \scalebox{0.75}{\fcolorbox{white}{white}{
      \begin{picture}(12,20) (34,-14)
        \SetWidth{2.5}
        \SetColor{Black}
        \Line(28,4)(52,4)
        \Line(52,-12)(28,-12)
        \SetWidth{1.0}
        \Line(48,4)(32,-12)
        \Line(32,4)(48,-12)
        \Line(40,-12)(40,4)
      \end{picture}
    }}
  \end{array} +
  \begin{array}{c}
    \scalebox{0.75}{\fcolorbox{white}{white}{
      \begin{picture}(12,20) (34,-14)
        \SetWidth{2.5}
        \SetColor{Black}
        \Line(28,4)(52,4)
        \Line(52,-12)(28,-12)
        \SetWidth{1.0}
        \Line(40,4)(32,-12)
        \Line(32,4)(48,-12)
        \Line(40,-12)(48,4)
      \end{picture}
    }}
  \end{array} +
  \begin{array}{c}
    \scalebox{0.75}{\fcolorbox{white}{white}{
      \begin{picture}(12,20) (34,-14)
        \SetWidth{2.5}
        \SetColor{Black}
        \Line(28,4)(52,4)
        \Line(52,-12)(28,-12)
        \SetWidth{1.0}
        \Line(48,4)(32,-12)
        \Line(40,4)(48,-12)
        \Line(40,-12)(32,4)
      \end{picture}
    }}
  \end{array} 
  + \text{\parbox{5em}{\centering (iteration terms)}} &=& 0
  \,,
\end{eqnarray*}
where the Feynman diagrams represent momentum-space expressions with
stripped-off colour factors.  Thick and thin lines represent static sources
and massless particles (scalars and gluons), respectively. In practice we can
ignore the contributions denoted by ``iteration terms'' since they are
generated by the logarithm of $W_C$ (see text below Eq.~(\ref{eq::WC})).  The
equations can be solved for the Feynman integrals involving pinches which one
in turn inserts into the expression for $V_o$ where they get multiplied by the
corresponding colour octet colour factor.

We have performed our calculation in general $R_\xi$ gauge and have 
checked that the final results for $V_s$ and $V_o$ are independent of 
$\xi$ which constitutes a welcome check.

As stated in Eq.~(\ref{eq::Eso}) it is necessary to add the ultrasoft
contribution to the potential in order to arrive at a finite quantity.
Actually the individual contributions are divergent and contain poles
in $\epsilon$. They cancel in the sum which is a strong check both for
$V_{s,o}$ and $\delta E^{\rm US}_{s,o}$.

\begin{figure}[t]
  \centering
  \begin{tabular}{c}
    \fcolorbox{white}{white}{
      \begin{picture}(146,46) (23,-41)
	\SetWidth{1.0}
	\SetColor{Black}
	\Arc[dash,dashsize=8](112,-5.727)(34.273,-159.017,-20.983)
	\Arc[dash,dashsize=8,clock](80,-30.273)(34.273,159.017,20.983)
	\Line[double,sep=3](24,-18)(80,-18)
	\SetWidth{3.0}
	\Line(80,-18)(112,-18)
	\SetWidth{1.0}
	\Line[double,sep=3](112,-18)(168,-18)
	\COval(80,-18)(4.243,4.243)(45.0){Black}{White}\Line(77.879,-20.121)(82.121,-15.879)\Line(77.879,-15.879)(82.121,-20.121)
	\COval(112,-18)(4.243,4.243)(45.0){Black}{White}\Line(109.879,-20.121)(114.121,-15.879)\Line(109.879,-15.879)(114.121,-20.121)
      \end{picture}
    }
  \end{tabular}
  \caption[]{\label{fig::us} 
    Feynman diagram which contributes to $\delta E^{\rm US}_{o}$.
    Single and double lines correspond to singlet and
    octet Greens functions, respectively. Dashed lines represent ultrasoft scalars.}
\end{figure}
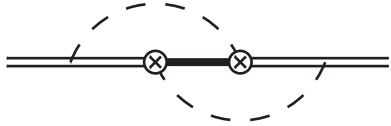

Using the results for $\delta E^{\rm US}_{s}$ from
Ref.~\cite{Stahlhofen:2012zx} we indeed arrive at a finite 
result for $E_s$. However, the poles do not cancel in the octet case.
After examining the calculation of $\delta E^{\rm US}_{o}$ we have
realized that the Feynman diagram in Fig.~\ref{fig::us} has not been
considered in~\cite{Stahlhofen:2012zx}. Only after taking it into account 
the result for $E_o$ becomes finite.
For completeness we provide the result of the missing contribution
which completes the list given in Appendix~B
of~\cite{Stahlhofen:2012zx}. Our result, which has been obtained in
$D=4-2\epsilon$ dimensions, reads
\begin{eqnarray}
  \mbox{(Fig.~\ref{fig::us}) = }
  \frac{i}{4\pi^D} g^4 \left(\frac{C_A}2 - C_F\right) \left(8C_F - 3C_A\right)
  \left(-\Delta V\right)^{2D-7} \frac{\Gamma^2\left(\frac
      D2-1\right)\Gamma(7-2D)}{(D-3)^2} 
  \,,
  \label{eq::us}
\end{eqnarray}
where $\Delta V = V_o - V_s$, $C_A=N_c, C_F=(N_c^2-1)/2/N_c$, and the coupling
$g$ is defined in Eq.~(\ref{eq::Lstat}).


\section{\label{sec::res}Results and conclusions}

In a first step we present results for the dimensionally regularized
potentials $V_s$ and $V_o$ which we parametrize in momentum space as
\begin{eqnarray}
  \tilde{V}_s &=& -\frac{8\pi C_F\alpha}{\vec{q}\,^2} \left[1 + \frac\alpha\pi
    \tilde{a}_s^{(1)} + \left(\frac\alpha\pi\right)^2 \tilde{a}_s^{(2)} +
    {\cal O}\left(\alpha^3\right)\right] \,,
  \nonumber \\ 
  \tilde{V}_o &=& -\frac{8\pi\left(C_F-\frac{C_A}2\right)\alpha}{\vec{q}\,^2}
  \left[1 + \frac\alpha\pi \tilde{a}_o^{(1)} + \left(\frac\alpha\pi\right)^2
    \tilde{a}_o^{(2)} + {\cal O}\left(\alpha^3\right)\right]\,, 
\end{eqnarray}
where $\alpha = g^2/4\pi$.
After adding all contributing diagrams we obtain for the colour singlet case
\begin{eqnarray}
  \tilde{a}_s^{(1)} &=& C_A\left[\frac1\epsilon  +
    \ln\left(\frac{4\pi\mu^2}{e^\gamma \vec{q}\,^2}\right) +
    \epsilon\left(\frac12\ln^2\left(\frac{4\pi\mu^2}{e^\gamma \vec{q}\,^2}\right) -
      \frac{\pi^2}{12}\right)\right] + {\cal O}(\epsilon^2)\,, \nonumber\\ 
  \tilde{a}_s^{(2)} &=& C_A^2\Bigg\{\frac1{2\epsilon^2}  + \left[\frac12 +
    \frac{\pi^2}6 + \ln\left(\frac{4\pi\mu^2}{e^\gamma
        \vec{q}\,^2}\right)\right]\frac1\epsilon \nonumber\\
  && \quad - 1 - \frac{\pi^2}{12} + \frac12\zeta(3) 
  + \left(1 + \frac{\pi^2}3\right) \ln\left(\frac{4\pi\mu^2}{e^\gamma \vec{q}\,^2}\right)
  + \ln^2\left(\frac{4\pi\mu^2}{e^\gamma \vec{q}\,^2}\right)\Bigg\}
  + {\cal O}(\epsilon)\,,
  \label{eq::a_mom1}
\end{eqnarray}
where $\gamma\approx0.57721\ldots$ is the Euler–Mascheroni constant.
The results for the colour-octet case read
\begin{eqnarray}
  \tilde{a}_o^{(1)} &=& \tilde{a}_s^{(1)}\,,\nonumber\\
  \tilde{a}_o^{(2)} &=& \tilde{a}_s^{(2)} + \delta \tilde{a}_o^{(2)}\,,\nonumber\\
  \delta \tilde{a}_o^{(2)} &=& -C_A^2 \pi^2 \left[\frac{1}{2\epsilon} +
    \ln\left(\frac{4\pi\mu^2}{e^\gamma \vec{q}\,^2}\right)\right] + {\cal
    O}(\epsilon)\text\,. 
  \label{eq::a_mom2}
\end{eqnarray}
One observes that, as for QCD, the one-loop results agree up to the change of
the global colour factor from $C_F$ to $C_F-C_A/2$ and that at two loops there is an
additional term proportional to $C_A^2\pi^2$. However, in contrast to QCD,
this term only contains a pole in $\epsilon$ and the corresponding logarithm.

In coordinate space we introduce $V_s$ and $V_o$ as
\begin{eqnarray}
  V_s &=& -\frac{2C_F\alpha}{r} \left[a_s^{(0)} + \frac\alpha\pi a_s^{(1)} +
    \left(\frac\alpha\pi\right)^2 a_s^{(2)} + {\cal
      O}\left(\alpha^3\right)\right]\,, \nonumber \\ 
  V_o &=& -\frac{2\left(C_F-\frac{C_A}2\right)\alpha}{r} \left[a_o^{(0)} +
    \frac\alpha\pi a_o^{(1)} + \left(\frac\alpha\pi\right)^2 a_o^{(2)} + {\cal
      O}\left(\alpha^3\right)\right]\,, 
  \label{eq::Vso}
\end{eqnarray}
and obtain for the coefficients
\begin{eqnarray}
  a_s^{(0)} &=& 1+ \epsilon\ln\left(4\pi \mu^2 e^\gamma r^2\right) +
  \epsilon^2\left(\frac{\pi^2}4 + \frac12\ln^2\left(4\pi \mu^2 e^\gamma
      r^2\right)\right) + {\cal O}\left(\epsilon^3\right)\,,\nonumber \\
  a_s^{(1)} &=& C_A\left[\frac1\epsilon + 2\ln\left(4\pi\mu^2 e^\gamma r^2\right) +
    \epsilon\left(\frac{5\pi^2}6 + 2\ln^2\left(4\pi\mu^2 e^\gamma
        r^2\right)\right)\right] + {\cal O}(\epsilon^2)\,,\nonumber\\ 
  a_s^{(2)} &=& C_A^2\bigg[\frac1{2\epsilon^2} + \left(\frac12 + \frac{\pi^2}6 
    + \frac32\ln\left(4\pi\mu^2 e^\gamma
      r^2\right)\right)\frac1\epsilon \nonumber\\ 
  &&\quad - 1 + \frac{7\pi^2}8  + \frac12\zeta(3) + \left(\frac32 +
      \frac{\pi^2}2 \right)  \ln\left(4\pi\mu^2 e^\gamma r^2\right) + \frac94
  \ln^2\left(4\pi\mu^2 
    e^\gamma r^2\right)\bigg] + {\cal O}(\epsilon)\,,\nonumber\\ 
  a_o^{(0)} &=& a_s^{(0)}\,,\nonumber\\
  a_o^{(1)} &=& a_s^{(1)}\,, \nonumber\\
  a_o^{(2)} &=& a_s^{(2)} + \delta a_o^{(2)}\,, \nonumber\\
  \delta a_o^{(2)} &=& -C_A^2 \pi^2 \left[\frac{1}{2\epsilon} + \frac{3}2
    \ln\left(4\pi\mu^2 e^\gamma r^2\right)\right] + {\cal O}(\epsilon)\,. 
  \label{eq::a_cor}
\end{eqnarray}
The ${\cal O}(\epsilon)$ terms in $a_s^{(0)}$ and $a_o^{(0)}$ arise due to the
$D$-dimensional Fourier transformation and are needed when inserting 
$\Delta V$ into the ultrasoft expression.

Note that there is no counterterm contribution since the
beta function vanishes in $\nfour$ SYM.
The poles in Eqs.~(\ref{eq::a_mom1}),~(\ref{eq::a_mom2}) and~(\ref{eq::a_cor}) 
are of infra-red type and cancel
against the ultraviolet poles of the ultrasoft contribution.
Within dimensional regularization such an identification is
necessary since scaleless integrals are set to zero (see, e.g.,
Ref.~\cite{Smirnov:2004ym}). 

For completeness we also display the ultrasoft result in coordinate space
which is given by~\cite{Pineda:2007kz,Stahlhofen:2012zx}
\begin{eqnarray}
  \delta E^{\rm US}_{s} &=& -\frac{2C_F\alpha}{r} \left[\frac\alpha\pi
    b_s^{(1)} + \left(\frac\alpha\pi\right)^2 b_s^{(2)} + {\cal
      O}\left(\alpha^3\right)\right]\,, \nonumber \\ 
  \delta E^{\rm US}_{o} &=& -\frac{2\left(C_F-\frac{C_A}2\right)\alpha}{r}
  \left[\frac\alpha\pi b_o^{(1)} + \left(\frac\alpha\pi\right)^2 b_o^{(2)}
    + {\cal O}\left(\alpha^3\right)\right]\,, 
  \label{eq::Eusso}
\end{eqnarray}
with
\begin{eqnarray}
  b_s^{(1)} &=& C_A\left[-\frac1\epsilon -2 + 2\ln\left(2C_A\alpha
      e^\gamma\right) - 2\ln\left(4\pi\mu^2 e^\gamma r^2\right)\right]\,, \nonumber\\ 
  b_s^{(2)} &=& C_A^2\bigg[-\frac1{2\epsilon^2} -\left(\frac12 + \frac{\pi^2}6
    + \frac32\ln\left(4\pi\mu^2 e^\gamma r^2\right)\right)\frac1\epsilon \nonumber\\
  &&\quad - \left(\frac32 + \frac{\pi^2}2\right)\ln\left(4\pi\mu^2 e^\gamma
    r^2\right)  - \frac94\ln^2\left(4\pi\mu^2 e^\gamma r^2\right) \nonumber\\ 
  &&\quad + \left(2 + \frac{2\pi^2}3\right) \ln\left(2C_A\alpha
    e^\gamma\right) + 2\ln^2\left(2C_A\alpha e^\gamma\right) \nonumber\\ 
  &&\quad - 6 - \frac{17\pi^2}{24} + 4\zeta(3)\bigg] - 2C_F C_A \frac{\pi^2}3\,,
  \nonumber\\ 
  b_o^{(1)} &=& b_s^{(1)}\,, \nonumber\\
  b_o^{(2)} &=& b_s^{(2)} + \delta b_o^{(2)}\,, \nonumber\\
  \delta b_o^{(2)} &=& C_A^2 \pi^2 \left[\frac{1}{2\epsilon} -
    2\ln\left(2C_A\alpha e^\gamma\right) + \frac{3}2\ln\left(4\pi\mu^2
      e^\gamma r^2\right)\right] 
  \,.
  \label{eq::b_us}
\end{eqnarray}
As compared to the result given in~\cite{Stahlhofen:2012zx} there is a change
in $b_o^{(2)} $ which is due to the missing Feynman diagram discussed in the
previous Section.
Note that initially the ultrasoft contribution depends on $\Delta V$
(cf. Eq.~(\ref{eq::us})) since $V_s$ and $V_o$ are present in the ultrasoft
singlet and octet propagators, respectively, see Appendix~A of
Ref.~\cite{Stahlhofen:2012zx}. 
In order to arrive at Eqs.~(\ref{eq::b_us}) the perturbative expansion of 
$\Delta V$ has been inserted and an expansion in $\alpha$ has been performed.

The comparison of the results in Eqs.~(\ref{eq::Vso}) and~(\ref{eq::Eusso}) shows
that in the sum the pole parts and the dependence on $\mu r$ cancels
and we arrive at the following results for the static energies
\begin{eqnarray}
  E_s &=& -\frac{2C_F\alpha}{r} \left[1 + \frac\alpha\pi e_s^{(1)} +
    \left(\frac\alpha\pi\right)^2 e_s^{(2)} + {\cal
      O}\left(\alpha^3\right)\right]\,, \nonumber\\ 
  E_o &=& -\frac{2\left(C_F-\frac{C_A}2\right)\alpha}{r} \left[1 +
    \frac\alpha\pi e_o^{(1)} + \left(\frac\alpha\pi\right)^2 e_o^{(2)} +
    {\cal O}\left(\alpha^3\right)\right]\,, 
  \label{eq::Eso2}
\end{eqnarray}
with
\begin{eqnarray}
  e_s^{(1)} &=& 2C_A\left[\ln\left(2C_A\alpha e^\gamma\right) - 1\right] \,,\nonumber\\
  e_s^{(2)} &=& 2C_A^2\left[\ln^2\left(2C_A\alpha e^\gamma\right) +
    \left(1+\frac{\pi^2}3\right)\ln\left(2C_A\alpha e^\gamma\right) +
    \frac{\pi^2}{12} - \frac72 + \frac94\zeta(3)\right] -
  2C_AC_F\frac{\pi^2}{3} \,,  \nonumber\\ 
  e_o^{(1)} &=& e_s^{(1)}\,, \nonumber \\
  e_o^{(2)} &=& e_s^{(2)} + \delta e_o^{(2)}\,, \nonumber\\
  \delta e_o^{(2)} &=& -2C_A^2 \pi^2 \ln\left(2C_A\alpha e^\gamma\right)
  \,.
  \label{eq::Eso2_res}
\end{eqnarray}
The result for $e_s^{(1)}$ and the quadratic logarithm in $e_s^{(2)}$
agree with Ref.~\cite{Pineda:2007kz} 
and the linear logarithm in $e_s^{(2)}$ coincides 
with~\cite{Stahlhofen:2012zx}.
It is interesting to note that the two-loop singlet and octet coefficients
only differ by a term proportional to $\pi^2$ multiplied by a logarithm which
originates from the ultrasoft contribution in Eq.~(\ref{eq::b_us}).

In the expressions for $e_o^{(1)}$ and $e_o^{(2)}$ as presented above only
the real part has been considered. As discussed in
Ref.~\cite{Stahlhofen:2012zx} there is a nonzero imaginary part in the
ultrasoft contribution which can be interpreted as the decay rate of the
octet state into the singlet state and massless particles. 
The result given in~\cite{Stahlhofen:2012zx} changes due 
to the additional Feynman diagram of Fig.~\ref{fig::us}
(cf. Eq.~(\ref{eq::us})). The corrected expression for 
$\Gamma_o = -2 \mbox{Im}[E_o(r)]$ reads 
\begin{eqnarray*}
  \Gamma_o &=& -\frac{8\alpha^2}r C_A\left(C_F-\frac{C_A}2\right)
  \left\{1+\frac\alpha\pi C_A\left[2\ln\left(2\alpha C_A e^\gamma\right) + 1 -
      \frac{2\pi^2}3\right]\right\}
  \,.
\end{eqnarray*}

The results discussed so far are obtained from a Wilson loop containing both
the coupling to the vector bosons and scalar fields of $\nfour$ SYM.
Alternatively, it is also possible to consider the ``ordinary'' Wilson loop
which is obtained by nullifying the term $\Phi_n|\dot{x}|$
in Eq.~(\ref{eq::WC}), i.e. there is no interaction of the static sources and
the scalars.  In analogy to Eq.~(\ref{eq::Vso}) we write
\begin{eqnarray}
  \bar{V}_s &=& -\frac{C_F\alpha}{r} \left[1 + \frac\alpha\pi \bar{a}_s^{(1)} +
    \left(\frac\alpha\pi\right)^2 \bar{a}_s^{(2)} + {\cal
      O}\left(\alpha^3\right)\right]\,, \nonumber \\ 
  \bar{V}_o &=& -\frac{\left(C_F-\frac{C_A}2\right)\alpha}{r} \left[1 +
    \frac\alpha\pi \bar{a}_o^{(1)} + \left(\frac\alpha\pi\right)^2 \bar{a}_o^{(2)} + {\cal
      O}\left(\alpha^3\right)\right]
  \label{eq::Vso_ord}
  \,.
\end{eqnarray}
The results for the coefficients read
\begin{eqnarray}
  \bar{a}_s^{(1)} &=& -C_A\,, \nonumber\\
  \bar{a}_s^{(2)} &=& C_A^2\left[\frac54 + \frac{\pi^2}4 - \frac{\pi^4}{64}\right]\,, \nonumber \\
  \bar{a}_o^{(1)} &=& \bar{a}_s^{(1)}\,, \nonumber \\
  \bar{a}_o^{(2)} &=& \bar{a}_s^{(2)} + \delta \bar{a}_o^{(2)}\,, \nonumber \\
  \delta \bar{a}_o^{(2)} &=& -C_A^2\left[\frac{3\pi^2}4 - \frac{\pi^4}{16}\right]
  \,.
\end{eqnarray}
As expected, these results are free from ultrasoft effects which 
contribute starting from three loops.
The expression for $\bar{a}_s^{(1)}$ agrees with the
literature~\cite{Alday:2007he,Pineda:2007kz},\footnote{There seems to be a
  misprint in the explicit result for $\bar{a}_s^{(1)}$ as given below
  Eq.~(31) of Ref.~\cite{Pineda:2007kz} since agreement
  with Ref.~\cite{Alday:2007he} is claimed. However, in~\cite{Alday:2007he} 
  a different sign between the tree-level and
  one-loop result is obtained.} the other three
results are new.

To conclude, we have computed two-loop corrections to the static potential
in $\nfour$ SYM. Together with previously computed ultrasoft contributions
two-loop expressions for the energy of two static sources are obtained where we
consider the latter both in a colour-singlet and colour-octet configuration.
Results are presented both for the ordinary Wilson loop and the one involving
only the interaction of the static sources and the vector field.



\section*{Acknowledgements}

We would like to thank Alexander Penin and Antonio Pineda for 
carefully reading the manuscript and for useful comments.
M.S. thanks Lance Dixon for drawing the attention to the topic of this paper.
This work was supported by DFG through SFB/TR~9.



\end{document}